\documentclass[12pt, preprint]{aastex6}

\usepackage{graphicx} 
\usepackage{bm} 
\usepackage{hyperref} 

\begin{document}


\title{
Search for Neutrinos in Super-Kamiokande associated with Gravitational Wave Events GW150914 and GW151226
}






\author{
K.~Abe\altaffilmark{1,32},
K.~Haga\altaffilmark{1},
Y.~Hayato\altaffilmark{1,32},
M.~Ikeda\altaffilmark{1},
K.~Iyogi\altaffilmark{1},
J.~Kameda\altaffilmark{1,32},
Y.~Kishimoto\altaffilmark{1,32},
M.~Miura\altaffilmark{1,32},
S.~Moriyama\altaffilmark{1,32},
M.~Nakahata\altaffilmark{1,32}
% ...
}
\altaffiltext{1}{Kamioka Observatory, Institute for Cosmic Ray Research, University of Tokyo, Kamioka, Gifu 506-1205, Japan}
% ...
\altaffiltext{32}{Kavli Institute for the Physics and Mathematics of the Universe (WPI), The University of Tokyo Institutes for Advanced Study, University of Tokyo, Kashiwa, Chiba 277-8583, Japan }
% ...

\collaboration{The Super-Kamiokande Collaboration}
\noaffiliation

\begin{abstract}
We report the results from a search in Super-Kamiokande for neutrino signals coincident with the first detected gravitational wave events, GW150914 and GW151226, using a neutrino energy range from 3.5~MeV to 100~PeV.
We searched for coincident neutrino events within a time window of $\pm$500 seconds around the gravitational wave detection time.
Four neutrino candidates are found for GW150914 and no candidates are found for GW151226.
The remaining neutrino candidates are consistent with the expected background events.
We calculated the 90\% confidence level upper limits on the combined neutrino fluence for both gravitational wave events, which depends on event energy and topologies.
Considering the upward going muon data set (1.6~GeV - 100~PeV) the neutrino fluence limit for each gravitational wave event is 14 - 37 (19 - 50) cm$^{-2}$ for muon neutrinos (muon antineutrinos), depending on the zenith angle of the event.
In the other data sets, the combined fluence limits for both gravitational wave events range from 2.4$\times 10^{4}$ to 7.0$\times 10^{9}$ cm$^{-2}$.
\end{abstract}

\renewcommand{\thefootnote}{\fnsymbol{footnote}}
\footnote[0]{$^{\dagger}$ Deceased.}
\footnote[0]{$^{\ddagger}$ Also at BMCC/CUNY, Science Department, New York, New York, USA.}

\maketitle


\section{Introduction} \label{sec:introduction}
On September 14th 2015 at 09:50:45 UT, LIGO identified the first evident signal of a gravitational wave as a coincidence of two chirp signals observed by two distant independent interferometers~\cite{ligo}.
It is suggested that the gravitational wave event, namely GW150914, was due to a merger of two black holes having masses of approximately 30 to 60 solar mass. 
After the announcement of the discovery by the LIGO team on February 11th 2016, many efforts to search for coincidences in the observational data in complementary experiments have been made~\cite{followups},
no astronomical counterpart has been identified except a weak coincident excess reported by Fermi GBM $\sim$0.4 s after GW159014~\cite{fermi} althogh still controversial result exists~\cite{fermi2}.
Because of the tremendous energies involved and the unknown nature of the region of the black hole merger, it is possible to imagine coincident production of neutrinos.
For example, the possibility of high energy neutrino emission from relativistic jets when an accretion disk is formed around the source has been discussed~\cite{highenu1, highenu2}.
IceCube and ANTARES have searched for high energy neutrinos above $\sim$100 GeV in a time window of $\pm$500 s around the real-time alert for GW150914 issued by LIGO, but reported no positive evidence for coincident neutrino events~\cite{iceant}.

Following this event the second gravitational wave signal, GW151226, was reported~\cite{ligo2}.
It was observed on 26th December 2015 at 03:38:53 UT, and was predicted to be due to a merger of two black holes having masses of 14.2 and 7.5 solar mass.
KamLAND reported no positive coincident neutrino events for both of these two gravitational wave events~\cite{kaml}.

We report the results of a search for neutrinos in coincidence with GW150914 and GW151226 in Super-Kamiokande (SK).
It is a water Cherenkov detector located 2700-meters-water-equivalent underground in Kamioka, Japan.
Detail of description of the detector can be found elsewhere~\cite{skdet}.
In this detector the Cherenkov ring pattern reconstruction identifies final state electron and muon direction and energy from which we infer the neutrino direction, flavor, and energy.
Our search includes the neutrino energy region of $1-100$~GeV, which is not covered by the previous searches with neutrino telescopes.
Moreover, SK has unique sensitivity to MeV neutrinos either from a core-collapse supernova in the Local Group or from some other similarly efficient mechanism.
The neutrino events with reconstructed energies above 100~MeV are categorized as the `high energy data sample' in SK and are typically used to study atmospheric neutrinos and search for proton decay.
Neutrino events with reconstructed energies down to 3.5~MeV are categorized as the `low energy data sample' and are typically used to study solar neutrinos and to search for core-collapse supernova neutrinos.
The background events in the search for astrophysical neutrinos for the high energy data sample are almost entirely atmospheric neutrinos, while radioactive impurities, spallation products from cosmic ray muons, atmospheric and solar neutrinos are the main backgrounds in the low energy data sample.

\section{Search method and results}\label{sec:searchmethod}
\subsection{High energy data sample}
The high energy data samples consist of three distinct topologies: fully-contained (FC), partially-contained (PC), and upward-going muon (UPMU). FC neutrino events have reconstructed interaction vertices inside the fiducial volume with little light detected in the outer detector. PC neutrino events also have interaction vertices within the fiducial volume of the inner detector, but there is significant light in the outer detector volume. UPMU neutrino events are the highest energy events in the SK detector and are due to muon neutrino interactions in the surrounding rock that produce penetrating muons. These muons either stop in the inner detector volume (stopping events) or continue through the inner detector (through-going events). All three data topologies are considered for this search. Further information about the event topologies, as well as the selection cuts used to identify them, can be found in \cite{ashie05}. 

A search window of $\pm$500 s around the LIGO detection time of each gravitational wave event is selected. This is consistent with the time window chosen by the authors of \cite{iceant}. Using data from 2339.4 days of livetime in SK, the number of neutrino events we expect to see in a 1000-second time window is (9.41$\pm$0.07)$\times$10$^{-2}$ for the FC data set, (7.52$\pm$0.23)$\times$10$^{-3}$ for the PC data set, and (1.65$\pm$0.03)$\times$10$^{-2}$ for the UPMU data set. In the search window around both GW150914 and GW151226, no neutrino events were found in the FC, PC, or UPMU data sets. This null result is used in the calculation of the upper limit on neutrino fluence in the subsequent sections of this paper.

\subsection{Low energy data sample}
There are two neutrino event selection algorithms for the low energy data sample: the supernova relic neutrino (SRN) search~\cite{sksrn} and the solar neutrino analysis~\cite{sk4sol}. 
The largest cross section in this energy region is the inverse beta decay of electron antineutrinos ($\bar{\nu}_e+p \rightarrow e^+ + n$). Neutrino elastic scattering ($\nu + e^{-} \rightarrow \nu + e^{-}$) is sensitive to all neutrino flavors, but dominated by electron neutrinos. The observable signal in the detector originates from the charged particle, i.e. positron or electron in these interactions. There are other charged-current and neutral-current interactions with $^{16}$O nuclei which are subdominant.

The target energy range for the SRN analysis is from 15.5~MeV to 79.5~MeV.
Backgrounds relevant to the SRN analysis are atmospheric neutrino interactions (decay electrons from invisible muons, neutral-current interactions, low-energy pions and muons), solar neutrino interactions and spallation products from cosmic ray muons.
The detailed analysis method and the reduction criteria are described in \cite{sksrn}, which, as well as this analysis, does not require a neutron capture in delayed coincidence~\cite{ntag}. 
After all the reduction steps, no neutino events are observed for the SRN analysis in the $\pm500$~s search window around both GW150914 and GW151226.
The expected number of background events in a 1000-second time window is $(1.45\pm0.09)\times 10^{-3}$ based on 1888.9 days of data. 
The absence of neutrino events in this energy region excludes the scenario of a coincidence between the gravitational wave and a core-collapse supernova within a distance of 260~kpc at 90\% C.L.

\begin{figure}[hptb]
\begin{center}
\includegraphics[width=15cm, bb=0 0 567 277]{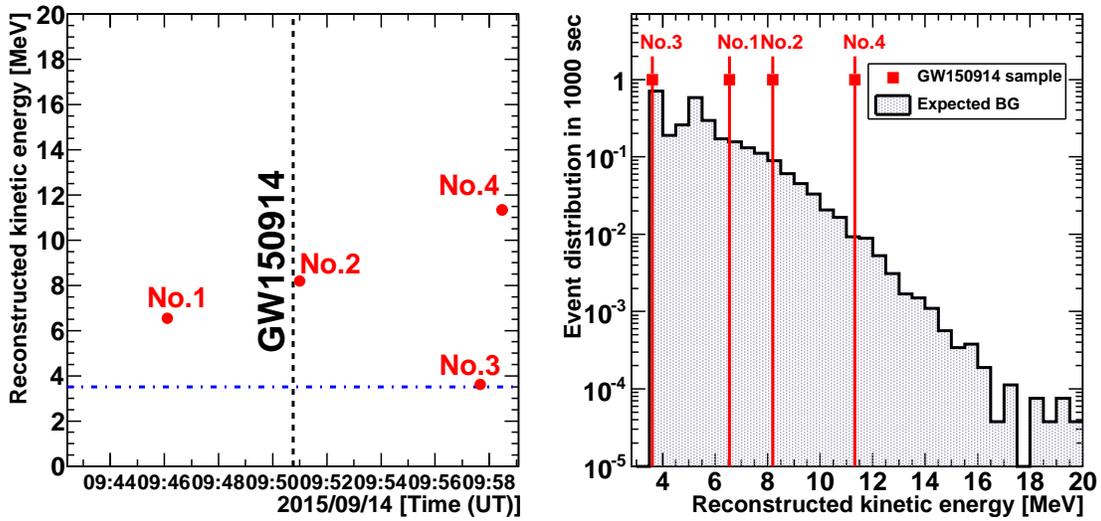}
\end{center}
\caption{Detection time and the observed kinetic energy of the charged particle (left) and the energy spectrum with the expected background spectrum (right) after reduction cuts for the solar neutrino analysis are applied to the 1000-second search window around GW150914.
The detection time of the four neutrino event candidates are as follows; No.1 9:46:07, No.2 9:51:00, No.3 9:57:40, No.4 9:58:28 UT.
The energy threshold of this analysis, shown with the blue dashed line, is 3.5~MeV in kinetic energy.
}
\label{solar_spect}
\end{figure}

The energy region considered in the solar neutrino analysis is lower than that of the SRN search, from 3.5~MeV to 19.5~MeV.
The fluence calculation will be applied below 15.5~MeV to avoid double counting with the SRN sample.
After applying all the reduction steps for the solar neutrino analysis, four neutrino event candidates remain within the $\pm500$ s search window around the LIGO detection time of GW150914, while no neutrino event candidates remain around GW151226.
Figure~\ref{solar_spect} shows the detection time and the energy of neutrino event candidates passing the reduction cuts for GW150914 (left) and the energy spectrum of the remaining events which is consistent with the background spectrum (right). 
On May 1st 2015, the trigger threshold was changed from 34 observed PMT signals within 200~ns to 31~hits~\cite{nakanod, yamada}.
Using the data after that (306.6 days of livetime) we expect $(2.90\pm0.01)$ events due to random coincidence for each 1000-second time window centered around a gravitational wave incident.
The probability of four or more events passing the reduction cuts is calculated to be 33.0\%, the probability to see none for GW151226 is 5.5\%.
Therefore, our findings are consistent with all of these neutrino event candidates being background.

The background events in the solar neutrino analysis consist mainly of radioactive impurities below 5.5~MeV and spallation products above 5.5~MeV. 
The first and second events are most likely remaining spallation events caused by $^{16}\mathrm{N}$~\cite{spabg}.
The transverse distance between each event and its preceding muon are 1.7~m and 1.2~m, the timing differences between the event and the preceding muon are 10.77 and 17.38 seconds, and the energies are above 6~MeV, making these consistent with spallation production.
Additionally, the likelihood value of the spallation product falls close to the cut value applied in the reduction steps.
The expected number of background events above 5.5~MeV that survive the reduction cuts is $(1.169\pm0.007)$; therefore the probability of two or more events remaining in our sample is 32.6\%.
The third event is most likely a radioactive background event caused by the beta decay of $\mathrm{^{214}Bi}$ from the radon decay chain. This is the dominant background near this energy region (3.6~MeV)~\cite{nakanod}.
The expected number of background events below 5.5~MeV that survive the reduction cuts is $(1.735\pm0.008)$; therefore the probability that one or more event remains in our sample is 82.4\%.
The fourth event is most likely a solar neutrino event because the reconstructed direction is close to the solar direction, (cosine of the angle between them ($\cos\theta_{\mathrm{sun}}$) is 0.96 with an angular resolution of 22.9~degrees) 
and its recoil electron kinetic energy is 11.3~MeV which is typical for solar neutrino event.
The main background of this energy region is spallation events, however, the chance of accidental coincidence of these background events with the solar direction is quite small since the reduction of spallation events are alredy fairly efficient.
The expected number of solar neutrino events is $(0.229\pm0.005)$ from the latest solar neutrino measurement in SK.
The probability of one or more solar events remaining in our sample is 20.5\%.

\begin{figure}
\begin{center}
\includegraphics[width=15cm, trim={0.5cm 3cm 0 1cm}, clip]{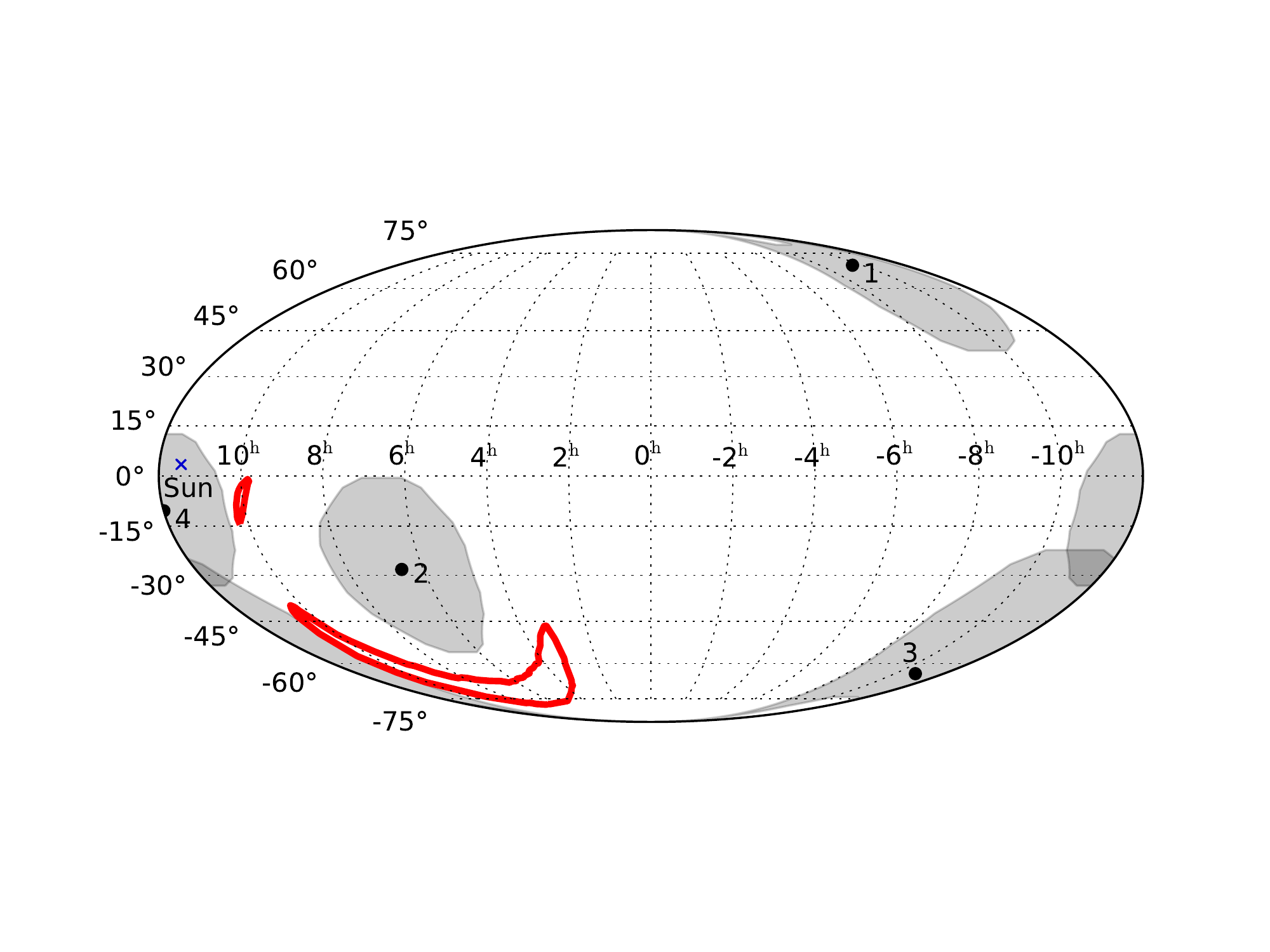}
\end{center}
\caption{Reconstructed directions of the charged particle in the remaining four events associated with GW150914 (black points), shown here with the 90 \% CL contour for the location of GW150914 according to LALInterference data \cite{veitch15}. The shaded area around SK events shows the 1$\sigma$ angular resolution of the charged particle which is calibrated by electron LINAC~\cite{linac} corresponding to the energy of each event. The cross shows the Sun's direction at the time of the GW event.}
\label{skymap}
\end{figure}

Figure~\ref{skymap} shows the sky map using the reconstructed direction of the charged particle in the remaining four events associated with GW150914.
Because the energy spectrum and species of the incoming neutrinos is not known, an estimate of the angular uncertainty on the direction of the incoming neutrino is difficult, however, the direction of the charged particle has a strong correlation with the incident neutrino direction for the case of neutrino-electron scattering, while a very weak anti-correlation exists in the case of inverse beta decay~\cite{vissani}. 

\subsection{LVT151012}
In addition to two gravitational wave signals LIGO reported a third binary black hole candidate with a smaller significance, named LVT151012, on October 12th 2015 at 09:54:43 UT~\cite{ligo2}. 
We also searched for a  coincident neutrino signal within a $\pm$ 500 s search window around the LVT151012 event.
No neutrino events are found in the FC, PC, UPMU and relic supernova neutrino data sample, while five neutrino event candidates remain after applying reduction steps for the solar neutrino analysis.
The probability of five or more events remaining in the solar neutrino data sample is calculated to be 16.8\% which is also consistent with background events.

\section{Neutrino fluence limit} \label{sec:fluence}
The number of neutrino candidate events observed in the search window can be converted to an upper limit on neutrino fluence for both of the gravitational wave
events. This is done separately for the low energy, FC+PC, and UPMU
data sets. The procedure used to extract the fluence limit will be
described briefly here, but follows the same prescription laid out in \cite{thrane09}. 

For the FC and PC data set, the neutrino fluence can be calculated
using equation (\ref{eqn:fluence}), 

\begin{equation} \label{eqn:fluence}
\Phi_{FC, PC} = \frac{N_{90}}{N_T \int dE_{\nu} \sigma (E_{\nu}) \epsilon (E_{\nu}) \lambda (E^{-2}_{\nu})},
\end{equation}

where $N_{90}$ is the 90\% C.L.
limit on the number of neutrino events in the search window calculated using a
Poisson distribution with a background, $N_{T}$ is the number of
target nuclei, $\sigma$ is the combined cross section for all
interactions particular to that neutrino flavor,
$\epsilon$ is the efficiency for measuring the neutrino event in the detector,
and $\lambda$ is the number density of the events for a given energy
spectrum with index of $-2$. This spectral index is commonly used for
astrophysical neutrinos accelerated by shocks \cite{gaisser94}. Since no neutrino candidate events pass cuts within the
search window for the FC and PC data sets, we derive an $N_{90}$ using
$N_{90} = - \ln(0.1) =$ 2.3. Fluence limits are calculated separately for each
neutrino species by considering the cross section and detection efficiency specific to the
neutrino type.

In this analysis, cross sections from NEUT 5.3.5 \cite{hayato09} are
used. To determine the detector efficiency, mono-energetic neutrino
interaction files produced with NEUT 5.3.5 are passed as inputs to the SK Monte Carlo detector simulation. 
The resulting output files then undergo the same reconstruction and
reduction procedures that are applied to the data. The efficiency is
determined by calculating the fraction of remaining events after
reduction and selection cuts. The energy range of 100 MeV - 10 GeV is considered. 

For the UPMU data set, the neutrino fluence is calculated using
equation (\ref{eqn:UPMUfluence})
 
\begin{equation} \label{eqn:UPMUfluence}
\Phi_{UPMU} = \frac{N_{90}}{A_{eff}(z)\int dE_{\nu} P(E_{\nu})S(z,E_{\nu})\lambda (E^{-2}_{\nu})}.
\end{equation}

Here, fluence depends on the zenith-dependent effective area
($A_{eff}(z)$) where $z$ is the zenith angle of the incoming neutrino, the probability for a
neutrino to create a muon with energy greater than $E_{\nu}^{min}$
($P(E_{\nu})$), the shadowing of the neutrinos due to interactions in
the Earth ($S(z,E_{\nu})$), and the number density ($\lambda$) of the
neutrino events for the given energy spectrum with index of $-2$. Again, since
there are no neutrino candidate events within the search window for the UPMU data set, we
use $N_{90}=$2.3. The energy range considered for the UPMU fluence is
1.6 GeV - 100 PeV. 

\begin{figure}[hptb]
\begin{center}
\includegraphics[width=8cm]{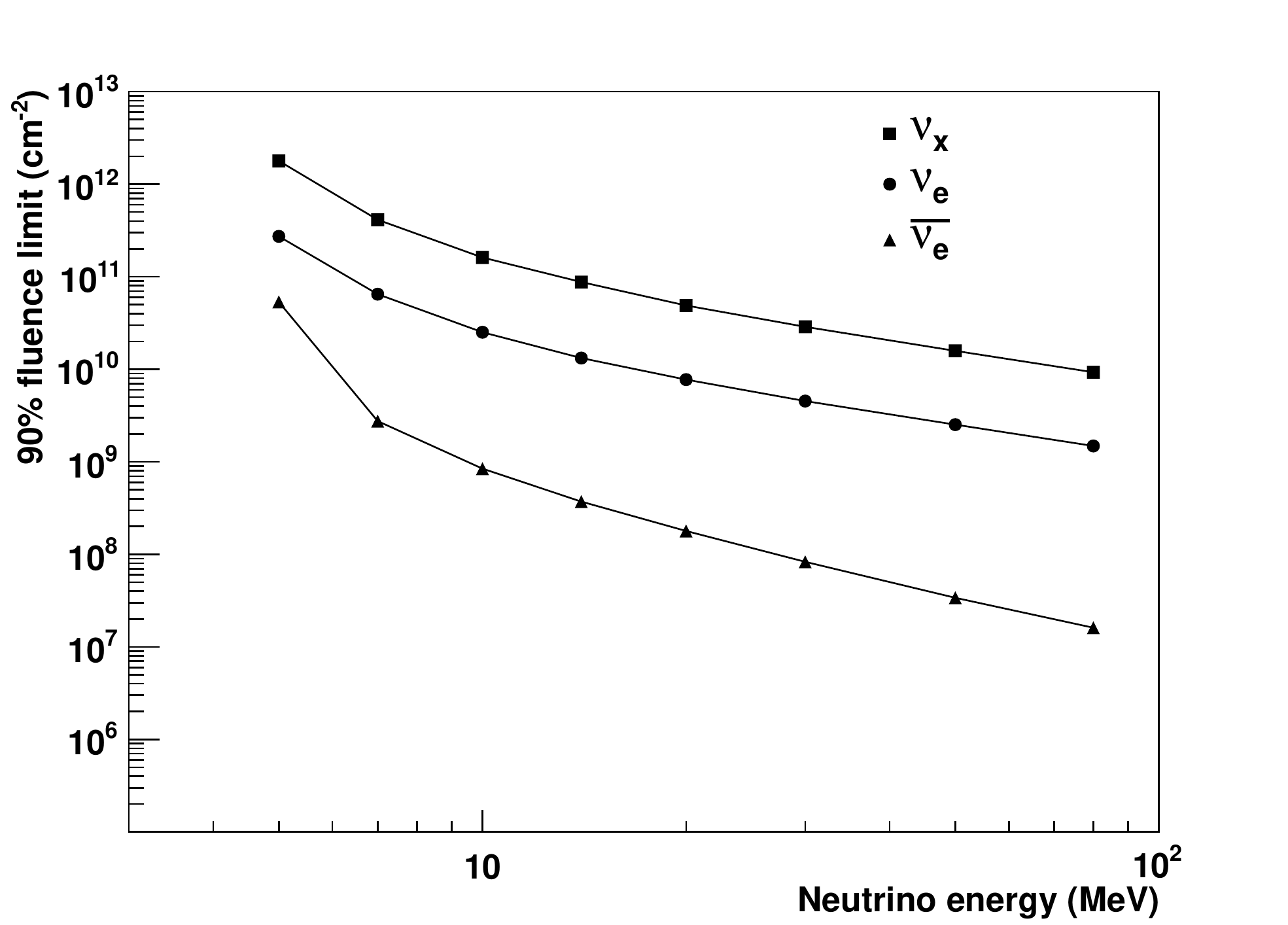}
\includegraphics[width=8cm]{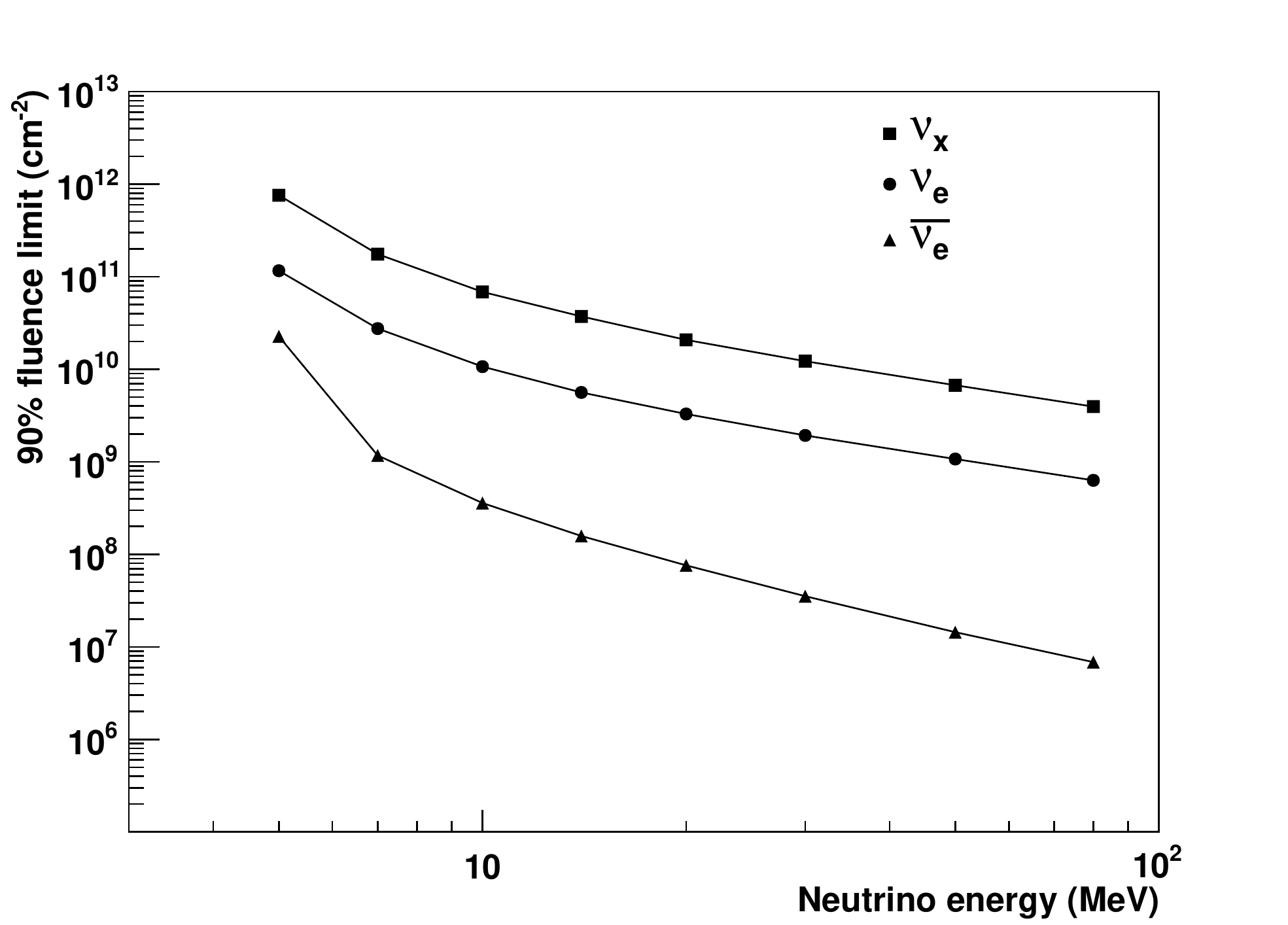}
\end{center}
\caption{The 90\% C.L. limit for GW150914 (left) and GW151226 (right) on fluence obtained by mono-energetic neutrinos at different specific energies from the low energy data sample.}
\label{fig:lowe_g}
\end{figure}

For the low energy data sample, the neutrino fluence calculation is
similar to equation (\ref{eqn:fluence}); however, since there is no
reason to assume a power spectrum nor any reliable theory of neutrino
emission spectrum in this energy region, we set the neutrino energy spectrum to have an index of $0$ which is flat.
Therefore, the fluence limit is calculated with the assumption of flat spectrum of neutrino energy from 3.5~MeV to 75~MeV as shown in equation (\ref{eqn:lowe}),

\begin{equation} \label{eqn:lowe}
\Phi_{lowe} = \frac{N_{90}}{N_T \int dE_{\nu} \lambda (E_{\nu}) \sigma (E_{\nu}) R(E_e, E_{vis}) \epsilon (E_{vis}) },
\end{equation}
where $R$ is the response function from electron or positron energy ($E_{e}$) to the kinetic energy in SK ($E_{vis}$). The response function and the detection efficiency ($\epsilon$) is calculated using the SK detector Monte Carlo simulation. 
$N_{90}$ is calculated to be 5.41 (2.30) from the remaining four (zero) neutrino candidate events compared to the expected 2.90 events in this data sample for GW150914 (GW151226).
The accidental coincidence of solar neutrinos with a gravitational wave signal are easily reduced by a factor of 0.9 with an angular cut on the recoil electron's direction ($\cos\theta_{\mathrm{sun}} > 0.8$).
In this case, the number of remaining neutrino candidate events is three (zero) compared to 2.46 expected events, and $N_{90}$ is 4.64 (2.30) for GW150914 (GW151226).
In addition, since any emission models for low energy neutrinos are not well motivated, we express the fluence limit which is calculated for monochromatic neutrino energy $E_{\nu}$ obtained by replacing $\lambda(E_{\nu})$ with a delta function $\delta(E-E_{\nu})$.
The fluence limits for both GW events are shown in Figure~\ref{fig:lowe_g}.

\begin{figure}[hptb]
\centering
\includegraphics[trim= 1cm 4cm 1cm 1cm, clip=true, width=0.49\textwidth]{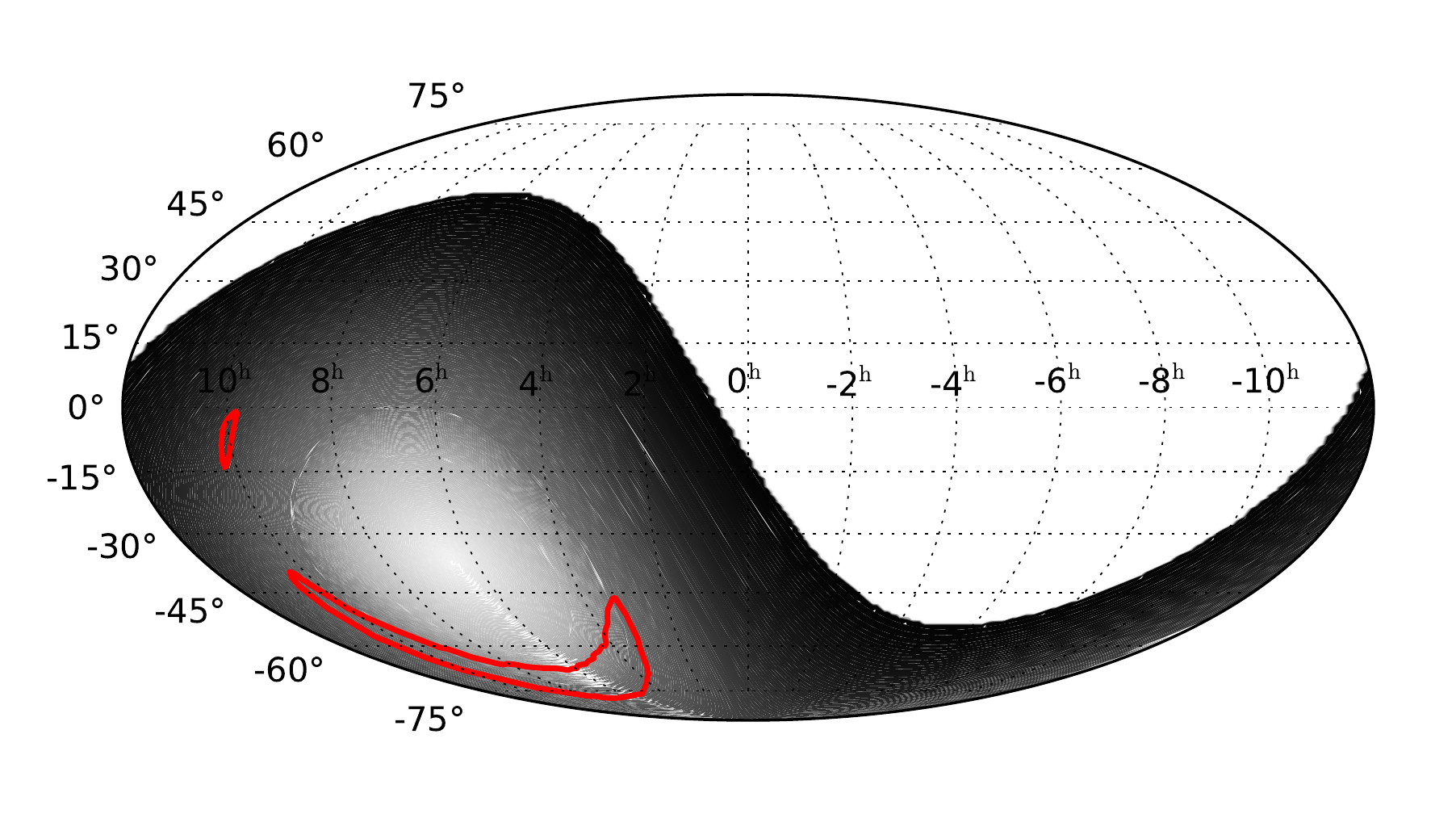}
\includegraphics[trim= 1cm 4cm 1cm 1cm, clip=true, width=0.49\textwidth]{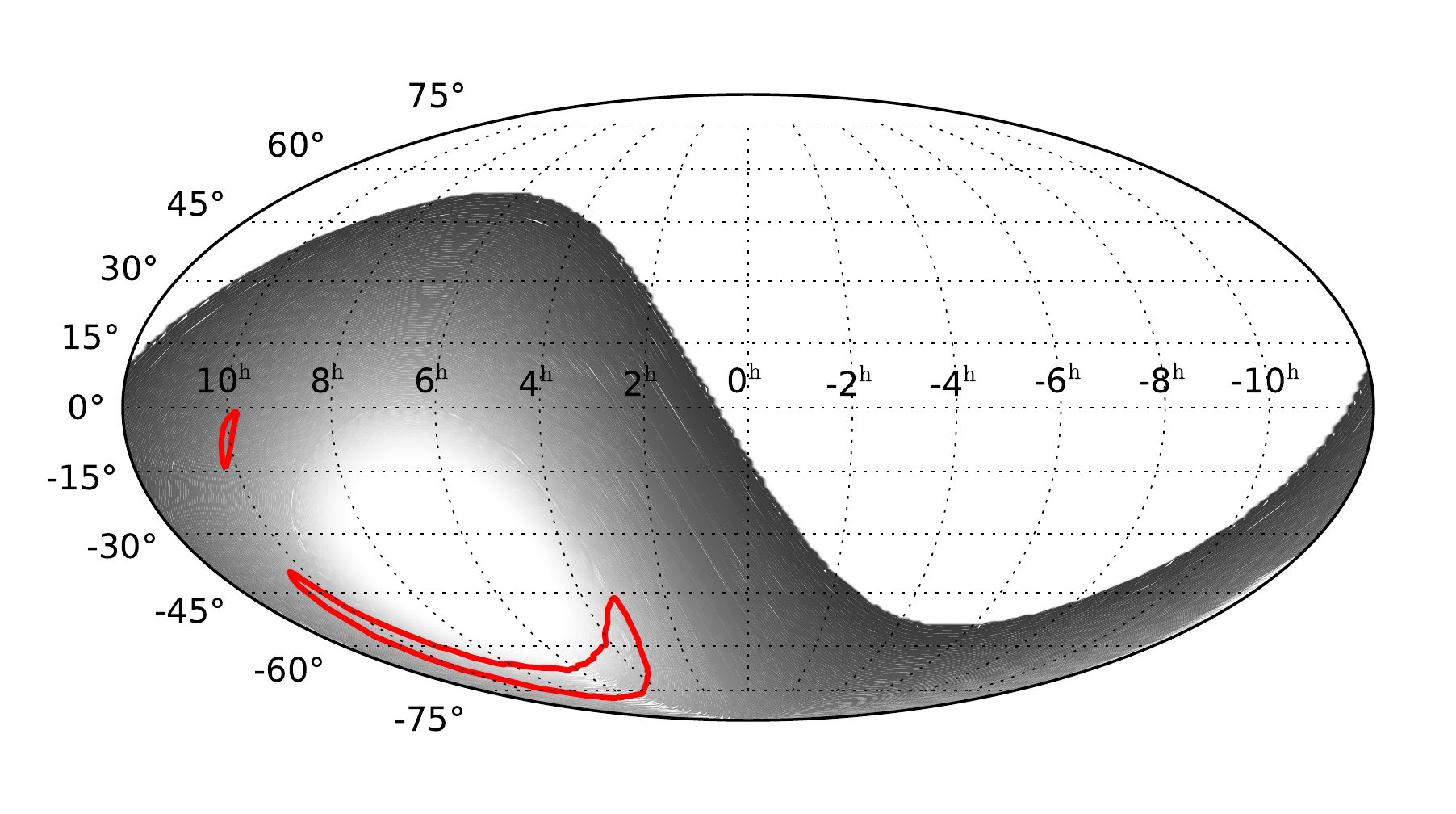} \\
\includegraphics[trim= 1cm 4cm 1cm 1cm, clip=true, width=0.49\textwidth]{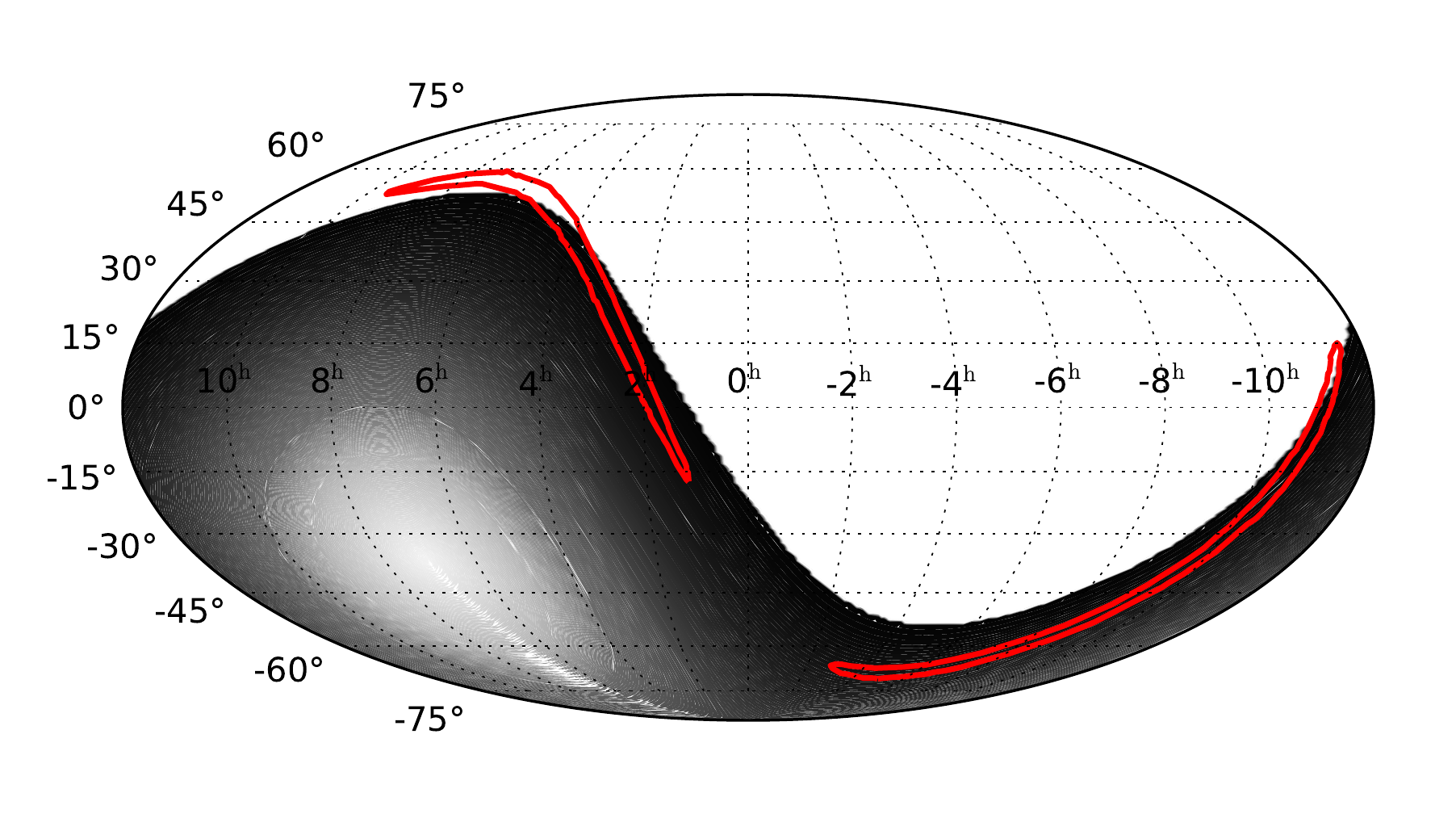}
\includegraphics[trim= 1cm 4cm 1cm 1cm, clip=true, width=0.49\textwidth]{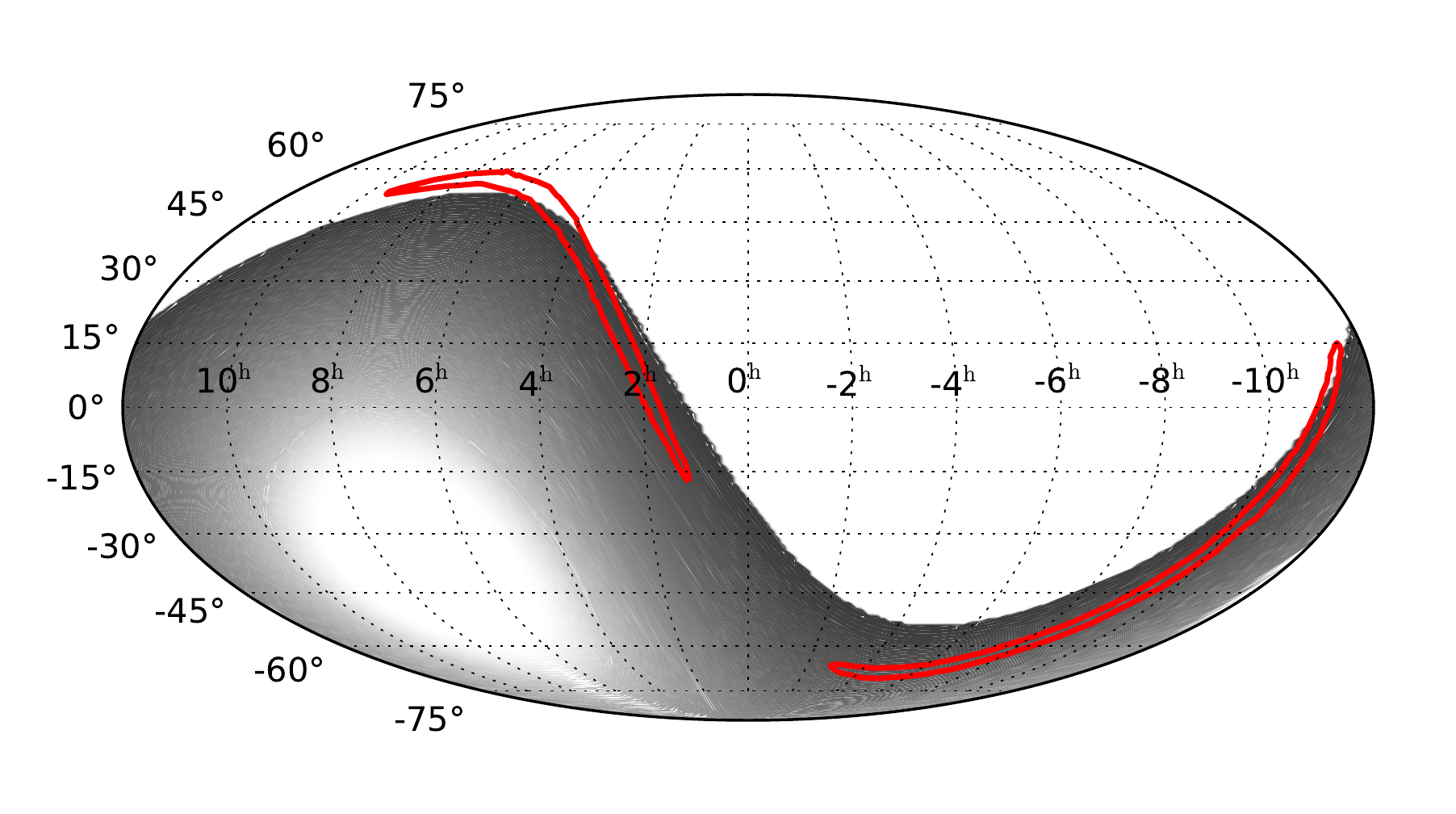}
\\
\includegraphics[trim= 2cm 1cm 2cm 11cm, clip=true, width=0.49\textwidth]{fig5.pdf}
\caption{The 90\% C.L. limit on fluence for neutrinos
considering only the UPMU data set, overlaid with the 90\% C.L.
contour for the location of the gravitational wave event according to LALInterference
 data \cite{veitch15}. The upper (lower) panel shows the information for
 GW150914 (GW151226), and the left (right) panel shows the information
 for neutrinos (antineutrinos). For this limit, an energy range of 1.6 GeV -100
 PeV was used.} \label{fig:UPMU_flu}
\end{figure}

Table~\ref{flu_table} shows the fluence limit results for the FC+PC, UPMU, and low energy data sets for both gravitational wave events separately, as well as for both events combined. To calculate the combined neutrino fluence limit, the $N_{90}$ is calculated using the search windows around both gravitational wave events and is then weighted by the number of gravitational wave events, which in this case is two. For the UPMU events, fluence is dependent on zenith angle, and thus we show the upper limit of neutrino fluence from UPMU events as a sky map of possible fluence limits in
Figure~\ref{fig:UPMU_flu}. Since the gravitational wave events are not well localized, a combined fluence limit for the UPMU data set is not reported. The UPMU upper limit fluence values range from ($14-37$) cm$^{-2}$ for neutrinos and from
($19-50$) cm$^{-2}$ for antineutrinos, depending on the direction of the
gravitational wave event. Converting our upper limit on fluence from
our UPMU sample to an upper limit on total energy radiated in
neutrinos by convolving equation \ref{eqn:UPMUfluence} by the energy
spectrum and weighting by $4 \pi  d_{GW}^2$ where $d_{GW}$ is the distance between the detector and the gravitational wave source. The resulting upper limit on total energy is $E_{\nu}^{\mathrm{tot}} \sim (1-6)\times 10^{55}$ ergs for
GW150914 assuming the LIGO-determined distance of 410 Mpc and
$E_{\nu}^{\mathrm{tot}} \sim (2-7)\times 10^{55}$ ergs for GW151226
assuming the LIGO-determined distance of 440 Mpc. For
comparison, the total energy radiated in gravitational waves by
GW150914 is $\sim 5 \times$10$^{54}$ ergs \cite{iceant} and the total
energy radiated in neutrinos from a typical supernova is $\sim$ a few
$\times$10$^{53}$ ergs \cite{bethe90}. 
\begin{table}[hptb]
\begin{center}
\begin{tabular}{lc c c}\hline \hline
 & GW150914 $\Phi_{\nu} $(cm$^{-2}$) & GW151226 $\Phi_{\nu}
$(cm$^{-2}$) & Combined $\Phi_{\nu} $(cm$^{-2}$) \\ \hline
 & {\bf from FC+PC only} \\
$\nu_{\mu}$       & $5.6 \times 10^4$ & $5.6 \times 10^4$ &  $2.8 \times 10^4$ \\
$\bar{\nu}_{\mu}$ &  $1.3 \times 10^5$ & $1.3 \times 10^5$ & $6.5 \times 10^4$\\
$\nu_e$           & $4.8 \times 10^4$ & $4.8 \times 10^4$ &  $2.4 \times 10^4$\\
$\bar{\nu}_e$     & $1.2 \times 10^5$ & $1.2 \times 10^5$ & $6.0 \times 10^4$ \\
\hline
 & {\bf from UPMU only} \\
$\nu_{\mu}$       & $14-37$ & $14-37$ & -  \\
$\bar{\nu}_{\mu}$ &  $19-50$ & $19-50$ & - \\
\hline
 & {\bf from low energy only} \\
$\bar{\nu}_e$     & 4.2 / 3.6 $\times 10^7$ & 1.8 / 1.8 $\times 10^7$ & 1.6 / 1.5 $\times 10^7$ \\
$\nu_e$           & 3.0 / 2.6 $\times 10^9$ & 1.3 / 1.3 $\times 10^9$ & 1.2 / 1.1 $\times 10^9$ \\
$\nu_x$           & 1.9 / 1.6 $\times 10^{10}$ & 8.1 / 8.1 $\times 10^9$ &  7.0 / 7.0 $\times 10^{9}$ \\
\hline
\end{tabular}
\caption{Limits at 90\% C.L. on the fluence of neutrinos from GW150914 and GW151226 given a spectral index of $\gamma=2$ and an energy range of 100 MeV - 10 GeV for the FC+PC and 1.6 GeV$-$100 PeV for the UPMU data samples, and a flat spectrum for the low energy data sample whose energy region is from 3.5 to 75~MeV. The values in left / right in lowe energy sample show the fluence limit without / with the solar direction cut, respectively.}
\label{flu_table}
\end{center}
\end{table}

\section{Conclusion}\label{sec:conclusion}
We search for possible neutrino signals coincident with GW150914 and GW151226 in the Super-Kamiokande detector using a wide energy range from 3.5~MeV to 100~PeV.
In the high energy data sample, three neutrino interaction categories are considered: FC, PC and UPMU.
No neutrino candidate events are found in the search window of $\pm$500 s around the LIGO detection time for both gravitational wave signals.

Low energy neutrino events are also examined using the SRN and the solar neutrino data samples in the same search time window.
No neutrino candidate events are found in the SRN data sample for both gravitational wave signals.
No neutrino candidate events are found in the solar neutrino data for GW151226; however four neutrino candidate events are found for GW150914.
These four events are well-understood as originating from radioactive background, spallation products from cosmic ray muons and solar neutrinos.

The obtained neutrino fluence limits give the most stringent limits for 
neutrino emission in the energy region below $\sim$100~GeV, assuming an $E^{-2}$ spectrum from GW150914 and GW151226.
The absence of MeV neutrino emission in the solar neutrino and the SRN data samples is inconsistent with the source of the gravitational wave signals being a near-by core-collapsed astronomical object.

\acknowledgments
We gratefully acknowledge the cooperation of the Kamioka Mining and Smelting Company.
The Super-Kamiokande experiment has been built and operated from funding by the Japanese Ministry of Education, Culture, Sports, Science and Technology, the U.S. Department of Energy, and the U.S. National Science Foundation. Some of us have been supported by funds from the Korean Research Foundation (BK21 and KNRC), the National Research Foundation of Korea (NRF-20110024009), the European Union (H2020 RISE-GA641540-SKPLUS) the Japan Society for the Promotion of Science, the National Natural Science Foundation of China under Grant No. 11235006, and the Scinet and Westgrid consortia of Compute Canada.


\begin{thebibliography}{}
\expandafter\ifx\csname natexlab\endcsname\relax\def\natexlab#1{#1}\fi

\bibitem[{{Abbott} {et~al.}(2016{\natexlab{a}}){Abbott}, {Abbott}, {Abbott},
  {Abernathy}, {Acernese}, {Ackley}, {Adams}, {Adams}, {Addesso}, {Adhikari},
  \& et~al.}]{ligo2}
{Abbott}, B.~P., {Abbott}, R., {Abbott}, T.~D., {et~al.} 2016{\natexlab{a}},
  Physical Review Letters, 116, 241103

\bibitem[{{Abbott} {et~al.}(2016{\natexlab{b}}){Abbott}, {Abbott}, {Abbott},
  {Abernathy}, {Acernese}, {Ackley}, {Adams}, {Adams}, {Addesso}, {Adhikari},
  \& et~al.}]{followups}
---. 2016{\natexlab{b}}, \apjl, 826, L13

\bibitem[{{Abbott} {et~al.}(2016{\natexlab{c}}){Abbott}, {Abbott}, {Abbott},
  {Abernathy}, {Acernese}, {Ackley}, {Adams}, {Adams}, {Addesso}, {Adhikari},
  \& et~al.}]{ligo}
---. 2016{\natexlab{c}}, Physical Review Letters, 116, 061102

\bibitem[{{Abe} {et~al.}(2016)}]{sk4sol}
{Abe}, K., {et~al.} 2016, ArXiv e-prints, arXiv:1606.07538

\bibitem[{{Adri{\'a}n-Mart{\'{\i}}nez}
  {et~al.}(2016){Adri{\'a}n-Mart{\'{\i}}nez}, {Albert}, {Andr{\'e}},
  {Anghinolfi}, {Anton}, {Ardid}, {Aubert}, {Avgitas}, {Baret},
  {Barrios-Mart{\'{\i}}}, \& et~al.}]{iceant}
{Adri{\'a}n-Mart{\'{\i}}nez}, S., {Albert}, A., {Andr{\'e}}, M., {et~al.} 2016,
  \prd, 93, 122010

\bibitem[{{Ashie} {et~al.}(2005)}]{ashie05}
{Ashie}, Y., {et~al.} 2005, \prd, 71, 112005

\bibitem[{{Bays} {et~al.}(2012){Bays}, {Iida}, {Abe}, {Hayato}, {Iyogi},
  {Kameda}, {Koshio}, {Marti}, {Miura}, {Moriyama}, {Nakahata}, {Nakayama},
  {Obayashi}, {Sekiya}, {Shiozawa}, {Suzuki}, {Takeda}, {Takenaga}, {Ueno},
  {Ueshima}, {Yamada}, {Yokozawa}, {Kaji}, {Kajita}, {Kaneyuki}, {McLachlan},
  {Okumura}, {Lee}, {Martens}, {Vagins}, {Labarga}, {Kearns}, {Litos}, {Raaf},
  {Stone}, {Sulak}, {Kropp}, {Mine}, {Regis}, {Renshaw}, {Smy}, {Sobel},
  {Ganezer}, {Hill}, {Keig}, {Cho}, {Jang}, {Kim}, {Lim}, {Albert},
  {Scholberg}, {Walter}, {Wendell}, {Wongjirad}, {Ishizuka}, {Tasaka},
  {Learned}, {Matsuno}, {Smith}, {Hasegawa}, {Ishida}, {Ishii}, {Kobayashi},
  {Nakadaira}, {Nakamura}, {Nishikawa}, {Oyama}, {Sakashita}, {Sekiguchi},
  {Tsukamoto}, {Suzuki}, {Takeuchi}, {Ikeda}, {Matsuoka}, {Minamino},
  {Murakami}, {Nakaya}, {Fukuda}, {Itow}, {Mitsuka}, {Miyake}, {Tanaka},
  {Hignight}, {Imber}, {Jung}, {Taylor}, {Yanagisawa}, {Kibayashi}, {Ishino},
  {Mino}, {Sakuda}, {Mori}, {Toyota}, {Kuno}, {Kim}, {Yang}, {Okazawa}, {Choi},
  {Nishijima}, {Koshiba}, {Totsuka}, {Yokoyama}, {Heng}, {Chen}, {Zhang},
  {Yang}, {Mijakowski}, {Connolly}, {Dziomba}, \& {Wilkes}}]{sksrn}
{Bays}, K., {Iida}, T., {Abe}, K., {et~al.} 2012, \prd, 85, 052007

\bibitem[{{Bethe}(1990)}]{bethe90}
{Bethe}, H.~A. 1990, Reviews of Modern Physics, 62, 801

\bibitem[{{Connaughton} {et~al.}(2016){Connaughton}, {Burns}, {Goldstein},
  {Blackburn}, {Briggs}, {Zhang}, {Camp}, {Christensen}, {Hui}, {Jenke},
  {Littenberg}, {McEnery}, {Racusin}, {Shawhan}, {Singer}, {Veitch},
  {Wilson-Hodge}, {Bhat}, {Bissaldi}, {Cleveland}, {Fitzpatrick}, {Giles},
  {Gibby}, {von Kienlin}, {Kippen}, {McBreen}, {Mailyan}, {Meegan}, {Paciesas},
  {Preece}, {Roberts}, {Sparke}, {Stanbro}, {Toelge}, \& {Veres}}]{fermi}
{Connaughton}, V., {Burns}, E., {Goldstein}, A., {et~al.} 2016, \apjl, 826, L6

\bibitem[{{Eichler} {et~al.}(1989){Eichler}, {Livio}, {Piran}, \&
  {Schramm}}]{highenu1}
{Eichler}, D., {Livio}, M., {Piran}, T., \& {Schramm}, D.~N. 1989, \nat, 340,
  126

\bibitem[{{Fukuda} {et~al.}(2003)}]{skdet}
{Fukuda}, S., {et~al.} 2003, Nuclear Instruments and Methods in Physics
  Research A, 501, 418

\bibitem[{Gaisser {et~al.}(1995)}]{gaisser94}
Gaisser, T.~K., {et~al.} 1995, Phys. Rept., 258, 173, [Erratum: Phys.
  Rept.271,355(1996)]

\bibitem[{{Gando} {et~al.}(2016)}]{kaml}
{Gando}, A., {et~al.} 2016, ArXiv e-prints, arXiv:1606.07155

\bibitem[{{Greiner} {et~al.}(2016){Greiner}, {Burgess}, {Savchenko}, \&
  {Yu}}]{fermi2}
{Greiner}, J., {Burgess}, J.~M., {Savchenko}, V., \& {Yu}, H.-F. 2016, ArXiv
  e-prints, arXiv:1606.00314

\bibitem[{Hayato(2009)}]{hayato09}
Hayato, Y. 2009, Acta Phys. Polon., B40, 2477

\bibitem[{{Nakahata} {et~al.}(1999)}]{linac}
{Nakahata}, M., {et~al.} 1999, Nuclear Instruments and Methods in Physics
  Research A, 421, 113

\bibitem[{Nakano(2016)}]{nakanod}
Nakano, Y. 2016, PhD thesis, University of Tokyo

\bibitem[{{Strumia} \& {Vissani}(2003)}]{vissani}
{Strumia}, A., \& {Vissani}, F. 2003, Physics Letters B, 564, 42

\bibitem[{Thrane {et~al.}(2009)}]{thrane09}
Thrane, E., {et~al.} 2009, \apj, 704, 503

\bibitem[{Veitch {et~al.}(2015)}]{veitch15}
Veitch, J., {et~al.} 2015, Phys. Rev. D, 91, 042003

\bibitem[{{Woosley}(1993)}]{highenu2}
{Woosley}, S.~E. 1993, \apj, 405, 273

\bibitem[{Yamada {et~al.}(2009)}]{yamada}
Yamada, S., {et~al.} 2009, IEEE Transactions on Nuclear Science, 57, 428

\bibitem[{{Zhang} {et~al.}(2015)}]{ntag}
{Zhang}, H., {et~al.} 2015, Astroparticle Physics, 60, 41

\bibitem[{{Zhang} {et~al.}(2016)}]{spabg}
{Zhang}, Y., {et~al.} 2016, \prd, 93, 012004

\end{thebibliography}

\end{document}